\documentclass[]{interact}

\usepackage{epstopdf}
\usepackage{subfigure}

\usepackage{natbib}
\bibpunct[, ]{(}{)}{;}{a}{}{,}
\renewcommand\bibfont{\fontsize{10}{12}\selectfont}

\usepackage{graphicx}
\usepackage{microtype}      
\usepackage{xcolor}
\usepackage{bm}
\usepackage{amsmath}
\usepackage{float}

\theoremstyle{plain}

\theoremstyle{definition}

\theoremstyle{remark}

\begin{document}


\title{Deep Learning for Market by Order Data}

\author{
\name{Zihao Zhang\textsuperscript{a}\thanks{CONTACT Zihao Zhang. Email: zihao@robots.ox.ac.uk}, Bryan Lim\textsuperscript{a} and Stefan Zohren \textsuperscript{a}}
\affil{\textsuperscript{a}Oxford-Man Institute of Quantitative Finance, University of Oxford, Oxford, UK.}
}

\maketitle

\begin{abstract}

Market by order (MBO) data -- a detailed feed of individual trade instructions for a given stock on an exchange -- is arguably one of the most granular sources of microstructure information. While limit order books (LOBs) are implicitly derived from it, MBO data is largely neglected by current academic literature which focuses primarily on LOB modelling. In this paper, we demonstrate the utility of MBO data for forecasting high-frequency price movements, providing an orthogonal source of information to LOB snapshots and expanding the universe of alpha discovery. We provide the first predictive analysis on MBO data by carefully introducing the data structure and presenting a specific normalisation scheme to consider level information in order books and to allow model training with multiple instruments. Through forecasting experiments using deep neural networks, we show that while MBO-driven and LOB-driven models individually provide similar performance, ensembles of the two can lead to improvements in forecasting accuracy -- indicating that MBO data is additive to LOB-based features.

\end{abstract}


\section{Introduction}
\label{introduction}

High-frequency microstructure data has received growing attention both in academia and industry with the computerisation of financial exchanges and the increase capacity of data storage. The detailed records of order flow and price dynamics provide us with a granular description of short-term supply and demand, and we can take the dynamics of order books into account during the modelling process. 
Propelled by the publication of the benchmark dataset \citep{ntakaris2018benchmark} of high-frequency limit order book (LOB) data, there has been a growing interest in research studying LOB data. Recent works by \cite{tsantekidis2017forecasting, sirignano2019universal, zhang2019deeplob, briola2020deep} demonstrate that strong predictive performance can be obtained from modelling high-frequency LOB data and with resulting predictions finding applications in market-making and trade execution which have short holding periods. 

In this work, we introduce Market by Order (MBO) data for predictive modelling with deep learning algorithms. MBO data provides full resolution of the underlying market microstructure -- with both LOB data and trade sequences being derived from it. 
Despite MBO data being the original raw data source, current literature on high-frequency predictive modelling focuses predominantly on LOBs and, to the best of our knowledge, MBO data has not been used for direct predicting modelling. We showcase that the usage of MBO data as an additional source information to LOB improves predictive performance and MBO data could inspire a range of meaningful features that are related to individual order positions.
 
A LOB is a record of all outstanding limit orders (passive orders) for an instrument at a given time point and it is sorted into different levels based on submitted prices. At each price level, a LOB only shows the total available quantity. However, any given price level actually consists of many individual orders with different sizes. MBO data is essentially a message-base data feed that allows us to infer the individual queue position for each individual order by reconstructing the order book step by step. A detailed description of MBO data and how it relates to LOB data is presented in Section~\ref{mbo}. 

We propose a deep learning model based on MBO data, and in particular, a classification framework is adopted to predict stock price movements. In doing so, we provide a complete analysis of MBO by carefully introducing the data structure and the components of the message-base data feed. A specific data normalisation scheme is introduced to model level information contained in LOBs and to allow model training with multiple instruments. Our dataset consists of MBO data over a period of one year for five highly liquid instruments from the London Stock Exchange. Our testing set contains millions of samples to verify the robustness and generalisation of the results.

In our proposed models, we apply deep learning architectures including LSTMs \citep{hochreiter1997long} and Attention mechanisms \citep{bahdanau2014neural} to model the dynamics of MBO data for market predictions. Our experiments show consistent and robust results from MBO data that are comparable to models that utilise derived LOB data. We observe that predictive models based on MBO data are complementary to LOB models and we propose an ensemble approach which yields superior results. As such, we observe that MBO data adds diversification to the LOB model and improves prediction performance. 

The remainder of the paper is organised as follows. After a short literature review in Section~\ref{literature}, we proceed in Section~\ref{mbo} by introducing MBO data, including data preprocessing, normalisation and labelling. Section~\ref{methodology} presents deep learning architectures. We next describe our experiments and present the results of predicting market movements from MBO data in Section~\ref{experiment}. We conclude our findings and discuss promising future extensions in Section~\ref{conclusion}.

\section{Literature}
\label{literature}

Research on the high-frequency microstructure data remains largely focused on modelling the limit order book (LOB), where the classical works are referred to \cite{o1995market, harris2003trading} and a review is presented in \cite{gould2013limit}. However, there is limited work on MBO data in the current literature. NASDAQ \citep{nasdaq} and CME Group \citep{cme} provide a preliminary description on MBO data for introducing their exchange match engines, and the works of \cite{byrd2019abides, belcak2020fast} use MBO data for market simulation to model trading scenarios or to study latency effects. To the best of our knowledge, this paper is the first to use MBO data to predict market movements, filling in this literature gap by using deep learning models.

Deep Learning \citep{goodfellow2016deep} algorithms have been heavily used for predicting high-frequency microstructure data \citep{tsantekidis2017forecasting, sirignano2019universal, briola2020deep, wallbridge2020transformers}.
In particular, \cite{zhang2018bdlob, zhang2019deeplob, zhang2019extending} apply convolutional neural networks and LSTMs to model the dynamics of LOB and demonstrate accuracy improvements over linear models. Unlike traditional time-series models \citep{mills1991time, hamilton2020time} or stochastic models \citep{islam2020comparison} that assume a parametric process for the underlying time-series, deep learning methods are able to capture arbitrary nonlinear relationships without placing any specific assumptions on the input data. Our experiments also suggest that deep networks deliver better results than linear methods for modelling MBO data. 

We investigate deep learning models, including LSTMs \citep{hochreiter1997long} and Attention \citep{bahdanau2014neural}, to model MBO data. Attention is used to solve the problem of diminishing performance with long input sequences by utilising information at each hidden state of a recurrent network \citep{bahdanau2014neural, dai2019transformer}, and it can be used for constructing multi-horizon forecasting models \citep{lim2020time}. Our experiment suggests that networks with a recurrent nature lead to good predictive results compared to the state-of-art networks trained with LOB data, suggesting the potential benefits of using MBO data as an additional data source.

\section{Market by Order Data}
\label{mbo}

\subsection{Descriptions of Market by Order Data}

In general, exchanges provide high-frequency microstructure data in three tiers, namely L1, L2 and L3, offering increasingly granular information and capabilities:
\begin{itemize}
\item Level 1 (L1):  L1 shows the price and quantity of the last executed trade and displays real time best bid and ask of an order book, also known as quote data;
\item Level 2 (L2): L2 data is more granular than L1 by showing bids and asks at deeper levels of an order book, and it is commonly referred as LOB data;
\item Level 3 (L3): L3 is essentially the MBO data introduced in this work and it provides even more information than L2 as it shows non-aggregated bids and asks placed by individual traders.   
\end{itemize}

In this work, we focus on MBO data, which is essentially a message-base data feed that allows us to observe individual actions of market participants. Essentially, it is an order instruction that describes the action of a specific trader at a given time point. In what follows, we focus on the essential components of such messages ignoring certain auxiliary information. Table~\ref{tb:mbo_example} shows an example of sequences of MBO data, where:
\begin{itemize}
\item \textbf{Time stamp} records the time point when an instruction is given;
\item \textbf{ID} shows the unique ID for order identification which is anonymous to others;
\item \textbf{Type} indicates the order type, here limit order (Type = 1) or market order (Type = 2);
\item \textbf{Side} indicates whether an order is buy (1) or sell (2);
\item \textbf{Action} represents the specific instruction where 0 means updating the price or size for the existing order, 1 means adding a new order and 2 means cancelling an existing order. If Action = 2, the entries of Side, Price and Size are N/A as the matching engine will be able to identify and cancel the existing order using the unique ID;
\item \textbf{Price} shows the price level of the instruction;
\item \textbf{Size} shows the size (i.e. number of stocks) of the instruction. 
\end{itemize}

\begin{table}[!t]
\tbl{An example of a sequence of market by order data.}
{\begin{tabular}{l|llllll}
\toprule
\textbf{Time stamp}                    & \textbf{ID}                 & \textbf{Type} & \textbf{Side} & \textbf{Action} & \textbf{Price} & \textbf{Size}   \\
\midrule
2018-01-02 09:21:15.717500766 & 462805645163273214 &1 & N/A    & 2      & N/A & N/A    \\
2018-01-02 09:21:18.585446702 & 462805645163298476 &1& 1    & 1      & 68.54 & 8334.0 \\
2018-01-02 09:21:20.680552032 & 462805645163297649 &1& 1    & 0      & 68.56 & 3227.0 \\
2018-01-02 09:21:20.944574722 & 462805645163297649 &1& N/A    & 2      & N/A & N/A    \\
2018-01-02 09:21:20.945483443 & 462805645163298567 &1& 2    & 1      & 68.59 & 5100.0       \\
\bottomrule
\end{tabular}}
\label{tb:mbo_example}
\end{table}

\begin{figure}[p]
\centering
\includegraphics[width=5.2in, height=4in]{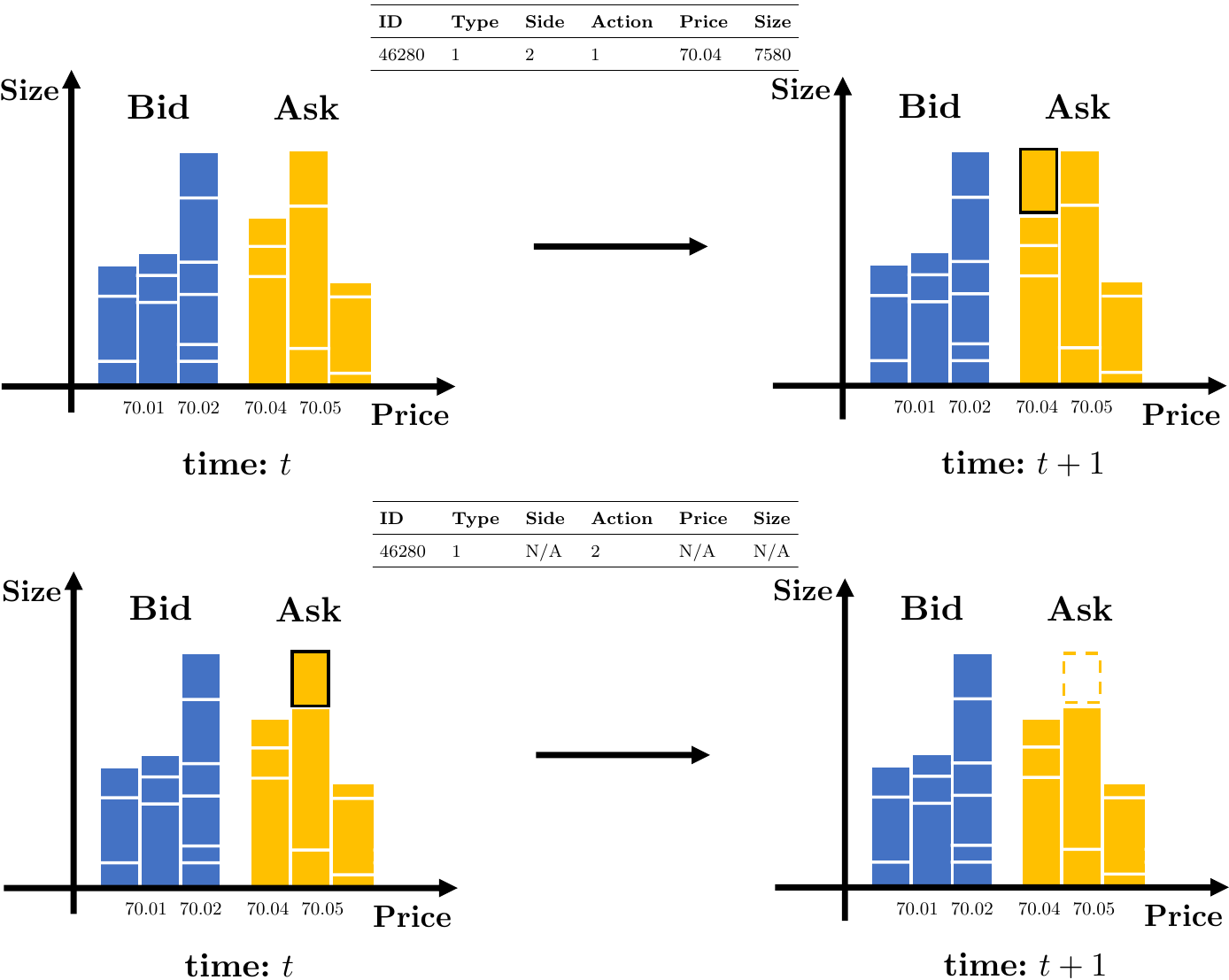}
\includegraphics[width=5.2in, height=4in]{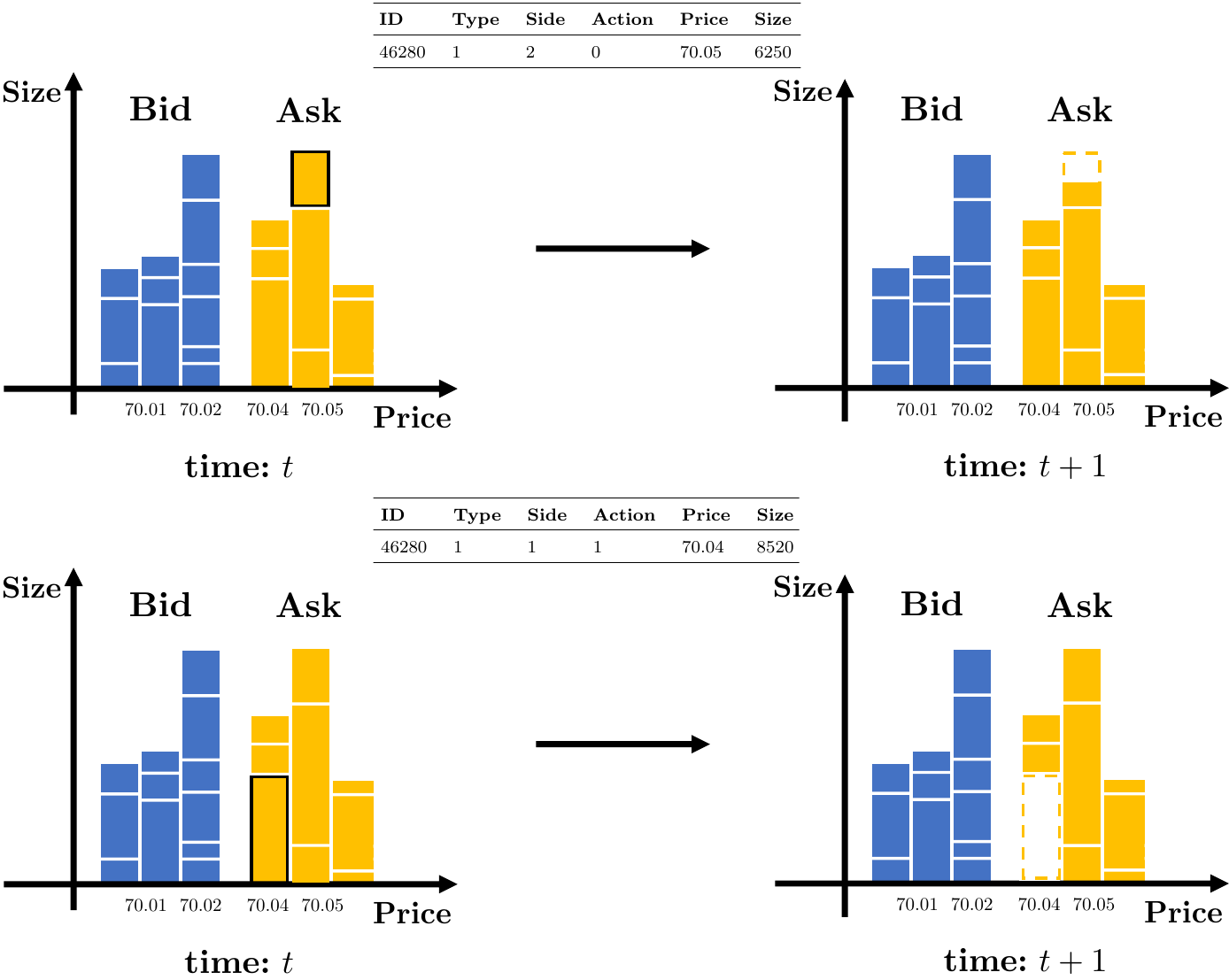}
\caption{An illustration of how MBO data updates a LOB. \textbf{Top:} An addition of a new limit order; \textbf{Middle top:} A cancellation of an existing order; \textbf{Middle bottom:} An update for a partial cancellation; \textbf{Bottom:} A marketable buy limit order that crosses the spread.}
\label{fig:mbo_mbp}
\end{figure}

A LOB updates whenever there is a new message from the MBO data coming in and this process is illustrated in Figure~\ref{fig:mbo_mbp}, where we show how a MBO message affect a LOB.
For example, if we look at the top of Figure~\ref{fig:mbo_mbp}, a new limit order (ID=46280) is added to the ask side of the order book with price at 70.04 and size of 7580. 
The order book updates its status and the new order is added to the right price level. In general, a LOB only shows the total available quantity at each price level but MBO data provides us with extra information by showing individual behaviour. Although, MBO data does not directly indicate which price level the order is added to, our normalisation scheme introduced in the next section allows us to consider this information and we not only obtain a smaller input space but also obtain relevant information comparable with LOB data. 

In addition, the usage of MBO data increases transparency and improves the understanding of order book dynamics without disclosing customer identification. Although, we can access to unique order ID but this number is generally assigned sequentially by the exchange match engine \citep{cme} and a private link is provided to the customer, which keeps identification confidential.
Further, unlike LOB data where we sometimes only view limited price levels, MBO data allows us to observe the entire order book with full-depth information. Such a granularity can improve traders' confidence in posting large order size as they can better evaluate the potential market impact by knowing individual queue positions.

\subsection{Data Preprocessing and Normalisation}

\begin{figure}[!t]
\centering
\includegraphics[width=5.5in, height=4.0in]{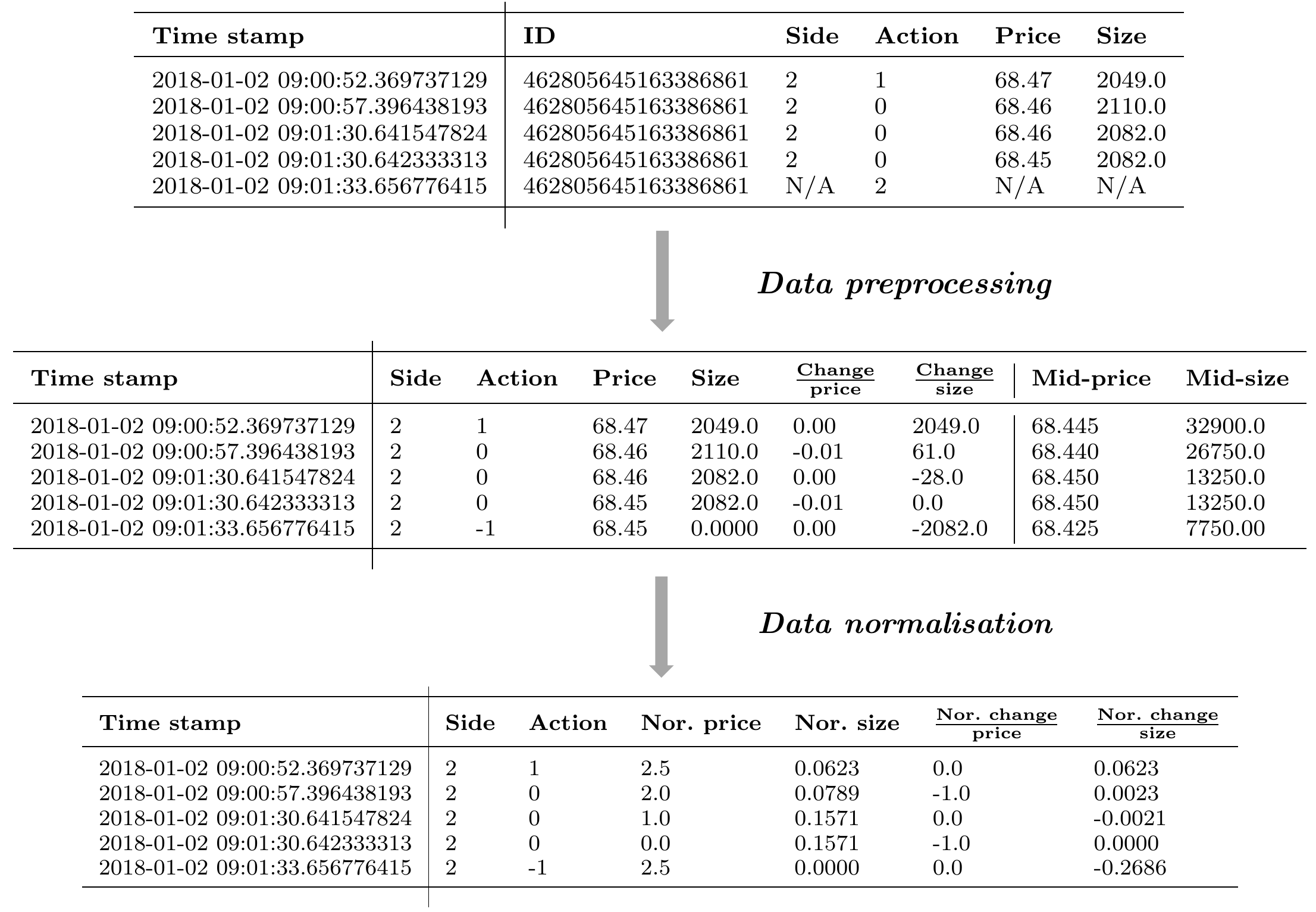}
\caption{An example of preprocessing and normalising the MBO data.}
\label{fig:mbo_normalisation}
\end{figure}

We focus on MBO data that represents limit orders because market orders only account for a tiny percentage of total order flow. 
Figure~\ref{fig:mbo_normalisation} illustrates the process of data preprocessing and normalisation. In particular, we process the MBO data for an unique ID as:
\begin{itemize}
\item \textbf{Side and Price:} Missing values correspond to updates and cancellations and we fill those with the corresponding values from the original order of that ID;
\item \textbf{Size:} Missing values correspond to full cancellation and we fill those with $0$ to indicate that no shares are outstanding after the action;
\item \textbf{Action:} we change Action to have values -1, 0 and 1. -1 means cancelling an order, 0 means updating price or size for the existing order and 1 means adding a new order;
\item \textbf{Change price} and \textbf{Change size}: We add these two new features to calculate the difference between entries for the price and size of a specific ID to reflect the intention of adding or decreasing positions for the given order.
\end{itemize} 
Data preprocessing is applied to every unique order ID and we then normalise the data as:
\begin{itemize}
\item \textbf{Normalised price}: (price - mid-price) / (minimum tick size $\times 100$).\footnote{Mid-price is the mean between the best ask and bid price. Those references prices and their sizes which we use below can be obtained on the fly from the MBO data.} This calculation transforms price to tick change, representing how many ticks the price is away from the mid-price. The deviation of minimum tick size is needed when we train models with multiple instruments as it maps price to a similar scale;
\item \textbf{Normalised size}: size / mid-size. Mid-size is the mean between the current best ask and bid size, which is similar to mid-price. 
\item \textbf{Normalised change price}: change price / minimum tick size;
\item \textbf{Normalised change size}: change size / mid-size.\footnote{By mid-size we denote the mean of best bid and ask size, in analogy to mid price. Those values again are obtained on the fly from the MBO data.}
\item \textbf{Side and Action}: remain unchanged.
\end{itemize} 

At the end, we remove ``Time stamp'' and ``ID'', leading to 6 features in our feature space. Note that the normalised price essentially represent the price level where the current order is in the order book, taking the level information into account. 

\subsection{Data Labelling}

In this work, we study a classification framework where we want to predict the future market movements into three classes: the market going up, staying stationary or going down. We use mid-prices to create labels and adopt the labelling method in \cite{zhang2019deeplob} to classify movements. In particular, we define
\begin{equation} \label{eq:label}
\begin{split}
&l_t = \frac{m_+(t) - m_-(t)}{m_-(t)}, \\
&m_-(t) = \frac{1}{k} \sum_{i=0}^{k-1} p_{t-i}, \\
&m_+(t) = \frac{1}{k} \sum_{i=1}^k p_{t+i},
\end{split}
\end{equation}
where $p_t$ is the mid-price at time $t$. We denote the prediction horizon as $k$ and it represents the number of arrivals of MBO data, meaning that we are working with tick time instead of clock time. To decide on the label we compare $l_t$ with a threshold ($\alpha$), labelling it as up if $l_t>\alpha$, down if $l_t<-\alpha$ and stationary otherwise. The choice of $\alpha$ is related to the prediction horizon ($k$) and we set $\alpha$ for each instrument to obtain a balanced training set. Our choices of $k$ and $\alpha$ are listed in Section~\ref{experiment} and we show that  the dataset are balanced under our choice. 

Note that Equation~\eqref{eq:label} introduces a smooth labelling that leads to consistent labels that are better for designing trading signals and the work of \cite{zhang2019deeplob} includes a more detailed discussion demonstrating the effects of different labelling methods. Interested readers are referred to their work for a detailed explanation.

\section{Methodology}
\label{methodology}

In this section, we introduce the different deep learning algorithms studied in our work. For a single input of any time-series, we write $\bm{x}_{1:T}$, where $\bm{x}_{t}$ represents the features at time $t$ and $T$ is the length of the sequence which will later correspond to the length of the lookback of the input.

\subsection{Multilayer perceptrons (MLPs)}
MLPs are canonical neural network models where a typical network is organised into a series of layers in a chain structure, with each layer being a function of the layer that precedes it. We can define the hidden layer of a MLP as: 
\begin{equation}
\bm{h}^{(l)} = g^{(l)} (\bm{W}^{(l)} \bm{h}^{(l-1)}  + \bm{b}^{(l)}          ),
\end{equation}
where $\bm{h}^{(l)} \in \mathbb{R}^{N_{l}}$ represents the $l$-th hidden layer with weights $\bm{W}^{(l)} \in \mathbb{R}^{N_l \times N_{l-1}}$ and biases $\bm{b}^{(l)} \in \mathbb{R}^{N_{l}}$. Here $g^{(l)}(\cdot)$ is the activation function that allows networks to model nonlinearities. The final output is a function of the last hidden layer and we compute objective functions to minimise errors between target outputs and estimates. 

However, for MLPs, we first need to flatten $\bm{x}_{1:T}$ and feed it to subsequent hidden layers. Doing this breaks the time dependences and treats features at different time stamps independently. We generally observe inferior results using MLPs and find that recurrent neural networks (RNNs) often deliver better performance. This is because a RNN acts as a memory buffer by summarising past information and recursively updating the hidden state with new observations at each time step of the input \citep{zhang2020deeppo}. 

\subsection{Long Short-Term Memory (LSTMs)}

Standard RNNs suffer from vanishing or exploding gradient problems \citep{bengio1994learning} and  Long Short-Term Memory networks (LSTMs) are proposed to solve this problem. This is done by operating a gating mechanism that efficiently controls the propagation of past information \citep{hochreiter1997long}. A LSTM updates its hidden state recursively and has a cell state $\bm{c}_t$ coupled with a series of gates at each hidden state. In mathematical terms, we can write 
\begin{equation}
\begin{split}
\text{Input gate:} \quad & \bm{i}_t = \sigma (\bm{W}_{i,h} \bm{h}_{t-1} +\bm{W}_{i,x}  \bm{x}_{t} + \bm{b}_i ), \\
&\text{with} \ \bm{W}_{i,h} \in \mathbb{R}^{N_h \times N_h},  \bm{W}_{i,x} \in \mathbb{R}^{N_h \times N_x} \ \text{and} \ \bm{b}_i \in \mathbb{R}^{N_h}, \\
\text{Output gate:} \quad & \bm{o}_t = \sigma (\bm{W}_{o,h} \bm{h}_{t-1} +\bm{W}_{o,x}  \bm{x}_{t} + \bm{b}_o ),\\
&\text{with} \ \bm{W}_{o,h} \in \mathbb{R}^{N_h \times N_h},  \bm{W}_{o,x} \in \mathbb{R}^{N_h \times N_x} \ \text{and} \ \bm{b}_o \in \mathbb{R}^{N_h}, \\
\text{Forget gate:} \quad & \bm{f}_t = \sigma (\bm{W}_{f,h} \bm{h}_{t-1} +\bm{W}_{f,x}  \bm{x}_{t} + \bm{b}_f),\\
&\text{with} \ \bm{W}_{f,h} \in \mathbb{R}^{N_h \times N_h},  \bm{W}_{f,x} \in \mathbb{R}^{N_h \times N_x} \ \text{and} \ \bm{b}_f \in \mathbb{R}^{N_h},
\end{split}
\end{equation} 
where $\bm{h}_{t-1}$ is the hidden state of a LSTM at time $t-1$ and $\sigma(\cdot)$ represents the sigmoid activation function. We use $\bm{W}$ and $\bm{b}$ to represent weights and biases at different gate operations. Subsequently, the current cell state and hidden state can be written as:
\begin{equation}
\begin{split}
\text{Cell state:} \quad & \bm{c}_t = \bm{f}_t \odot \bm{c}_{t-1} + \bm{i}_t \odot \text{tanh} (\bm{W}_{c,h} \bm{h}_{t-1} +\bm{W}_{c,x}  \bm{x}_{t} + \bm{b}_c), \\
\text{Hidden state:}: \quad & \bm{h}_t = \bm{o}_t \odot \text{tanh} (\bm{c}_t), 
\end{split}
\end{equation}
where $\bm{W}_{c,h} \in \mathbb{R}^{N_h \times N_h}$, $\bm{W}_{c,x} \in \mathbb{R}^{N_h \times N_x}$, $\bm{b}_c \in \mathbb{R}^{N_h}$, $\odot$ is the element-wise product and $\text{tanh}(\cdot)$ is the hyperbolic tangent activation function. The hidden state $\bm{h}_t$ summarises the information from past states and current observations, and the gating mechanism efficiently addresses the vanishing gradient problem.

\subsection{Attention Mechanism}

The Attention Mechanism \citep{bahdanau2014neural} is heavily used in machine translation and is proposed to solve the problem of diminishing performance for long input sequences. On the one hand, a LSTM calculates the final output as a function of only the last hidden state. An attention model, on the other hand, with an additional component called context vector, assigns trainable weights to all the hidden states of an input. We can write an attention mechanism for modelling many-to-one problem as:
\begin{equation}
\bm{h}_{t} = \bm{f}_t(\bm{h}_{t-1}, \bm{x}_{t}),
\end{equation}
where $\bm{h}_{t}$ can be the hidden state from a LSTM at time $t$ for an input $\bm{x}_{1:T}$, and we define the context vector $\bm{c}_{T}$ as:
\begin{equation}
  \begin{alignedat}{2}
    & \text{Convext vector:} &&\bm{c}_{T} = \sum_{t=1}^{T} \alpha(t, T)\bm{h}_{t}, \\
    & \text{Attention weights:} \quad &&\alpha(t, T)  = \frac{exp(e(t,T))}{\sum_{t=1}^{T} exp(e(t,T))}, \\
    &\text{Score:} && e(t,T) = \bm{v}^T \text{tanh}(\bm{W}_h \bm{h}_{t}),
  \end{alignedat}
\end{equation}
where $\bm{v} \in \mathbb{R}^{N_h}$ and $\bm{W}_h \in \mathbb{R}^{N_h \times N_h}$ are the trainable weights. We can then obtain the attention vector:
\begin{equation}
\bm{a}_T = \bm{f}(\bm{c}_T, \bm{h}_T) = \text{tanh}(\bm{W}_c[\bm{c}_T; \bm{h}_T]),
\end{equation}
where the final output is a function of $\bm{c}_{T}$, taking information at every hidden state into account.

\section{Experiments}
\label{experiment}

\subsection{Descriptions of Datasets}

Our datasets consist of MBO data for five highly liquid stocks, Lloyds (LLOY), Barclays (BARC), Tesco (TSCO), BT and Vodafone (VOD), for the entire year of 2018 from the London Stock Exchange. From the MBO data one can derive LOB data which we use for our benchmarks and for references prices. Our LOB dataset contains ask and bid information for an order book up to ten levels. For our modelling we remove messages outside ten levels from the MBO data to align the timestamps of two datasets allowing for fair comparisons in the performance analysis. Afterwards, we train two sets of models by separately using the MBO and LOB data with the same targets. A direct comparison can be then made to compare predictive performance using the MBO and LOB data respectively. 

For each trading day, we take the data between 08:30:00 and 16:00:00, restricting ourselves to liquid continuous trading hours, excluding any auctions. Overall, we have more than 169 million samples in our dataset and we take the first 6 months as training data, the next 3 months as validation data and the last 3 months as testing data. In the context of high-frequency microstructure data, we have more than 46 million observations in our testing set, providing sufficient scope for verifying the robustness and generalisability of model performance.  

We test our models at three prediction horizons ($k = 20, 50, 100$) and list the choices of label parameter ($\alpha$) in Table~\ref{tb:setting}. We choose $\alpha$ for each instrument to have a balanced training set and the proportion of different classes is presented in Figure~\ref{fig:label_class} in Appendix~\ref{add_results}. Overall, the labels are roughly balanced for the testing set as well (noting that those were fixed on the training set). In terms of the lookback window ($T$) of the input, we take the 50 most recent updates of MBO data to form a single input and feed it to our model. 
Note that we are working with tick time instead of physical clock time. In other words, the notation of time step refers to the arrival of MBO updates. One advantage of working with tick time is to deal with uneven trading volumes throughout a day. When a market opens with great volatility, we obtain more ticks and the model naturally makes faster predictions.

\begin{table}[!t]
\tbl{Label parameters ($\alpha$) for different prediction horizons and instruments (units in $10^{-4}$).}
{\begin{tabular}{l|lllll}
\toprule
        & LLOY & BARC & TSCO & BT   & VOD  \\
\midrule
k = 20  & 0.25 & 0.35 & 0.10 & 0.40 & 0.22 \\
k = 50  & 0.50 & 0.65 & 0.70 & 0.70 & 0.45 \\
k = 100 & 0.75 & 0.95 & 1.20 & 1.00 & 0.70 \\
\bottomrule
\end{tabular}}
\label{tb:setting}
\end{table}

\subsection{Training Procedure}

For the MBO data, we study the deep learning models (MBO-MLP, MBO-LSTM and MBO-Attention) introduced in Section~\ref{methodology} along with a simple linear model (MBO-LM). We list the values of hyperparameters for different algorithms in Table~\ref{tb:model_setting}, and the Gradient descent with the Adam optimiser \citep{kingma2014adam} is used for training all models. The complete search space of hyperparameters is included in Appendix~\ref{search_space} and we use a grid-search method to select best hyperparameters. 

For the LOB data, we include the 10 levels of a limit order book and past 50 observations as a single input. We follow the normalisation scheme in \cite{zhang2019deeplob} and both the MBO and LOB datasets share the same predictive targets, allowing a direct comparison between different models. 
We choose state-of-art network architectures as comparison models, including the LOB-LSTM \citep{sirignano2019universal}, LOB-CNN \citep{tsantekidis2017forecasting} and LOB-DeepLOB \citep{zhang2019deeplob}. The details of the network architecture and choices of hyperparameters can be found in their papers. We also include a linear model (LOB-LM) and a multilayer perception (LOB-MLP) as benchmark models. 

We use categorical cross-entropy loss as our objective function and the learning is stopped when the validation loss does not decrease for more than 10 epochs. In general, it takes about 30 epochs to finish model training. TensorFlow and Keras \citep{girija2016tensorflow} are used to build all models and four NVIDIA GeForce RTX 2080 are used in our experiment.

\begin{table}[H]
\tbl{Choices of hyperparameters.}
{\begin{tabular}{l|lllllll}
\toprule
          & \textbf{$\frac{\text{\# of}}{\text{layers}}$} & \textbf{$\frac{\text{\# of}}{\text{units}}$} & \textbf{$\frac{\text{Learning}}{\text{rate}}$} & \textbf{$\frac{\text{Batch}}{\text{size}}$}  & \textbf{$\frac{\text{\# of}}{\text{parameters}}$} \\
\midrule
LM       & -            & -                 & 0.0001        & 128                              & 903            \\
MLP       & 1            & 64                 & 0.0001        & 128                               & 19459            \\
LSTM      & 2            & 64                 & 0.0001        & 128                              & 51907            \\
Attention & 2            & 64                 & 0.0001        & 128              & 72067            \\
\bottomrule
\end{tabular}}
\label{tb:model_setting}
\end{table}

\subsection{Experimental Results}

Table~\ref{tb:results} summarises the results for all models studied (different rows) and one suitable for each different prediction horizons. We use four evaluation metrics (different columns) to make comparisons: Accuracy, Precision, Recall and F1-score. Kolmogorov-Smirnov \citep{massey1951kolmogorov} tests are used to check the statistical significance of results and all differences in evaluation metrics are significant.

\begin{table}[!t]
\tbl{Experimental results for different prediction horizons ($k$).}
{\begin{tabular}{l|llll}
\toprule
\textbf{Model} & \textbf{Accuracy \%} & \textbf{Precision \%} & \textbf{Recall \%} & \textbf{F1 \%} \\
\midrule
\multicolumn{5}{c}{\textbf{Prediction horizon k = 20}}                         \\
\midrule
MBO-LM                   & 41.81       & 41.16        & 41.81     & 34.97 \\
MBO-MLP                  & 47.12       & 46.17        & 47.12     & 46.46 \\
MBO-LSTM                 & 61.94       & 61.60        & 61.94     & 61.75 \\
MBO-Attention            & 61.19       & 62.83        & 61.19     & 61.73 \\
\midrule
LOB-LM                   & 45.71       & 43.44        & 45.71     & 42.38 \\
LOB-MLP      & 50.06       & 50.04        & 50.06     & 46.89 \\
LOB-LSTM        & 66.09       & 67.53        & 66.09     & 66.68 \\
LOB-CNN  & 63.39       & 67.31        & 63.39     & 64.64 \\
LOB-DeepLOB     & 68.73       & 68.16        & 68.73     & 68.40\\
\midrule
Ensemble-MBO     & 62.35       & 62.92        & 62.35     & 62.56\\
Ensemble-LOB     & 67.97       & 68.74        & 67.97     & 68.31\\
Ensemble-MBO-LOB     & \textbf{68.95}       & \textbf{69.10}        & \textbf{68.95}     & \textbf{69.02}\\
\midrule
\multicolumn{5}{c}{\textbf{Prediction horizon k = 50}}                  \\
\midrule
MBO-LM        & 41.88       & 38.42        & 41.88     & 36.57 \\
MBO-MLP       & 46.39       & 43.07        & 46.39     & 42.33 \\
MBO-LSTM      & 58.84       & 59.65        & 58.84     & 59.18 \\
MBO-Attention & 59.31       & 56.10        & 59.31     & 56.88 \\
\midrule
LOB-LM        & 46.97       & 44.34        & 46.97     & 41.13 \\
LOB-MLP       & 50.56       & 48.46        & 50.56     & 47.25 \\
LOB-LSTM      & 64.49       & \textbf{64.88}        & 64.49     & 64.65 \\
LOB-CNN       & 64.77       & 62.55        & 64.77     & 63.26 \\
LOB-DeepLOB   & 66.12       & 64.37        & 65.38     & 64.79\\
\midrule
Ensemble-MBO     & 60.03       & 58.45        & 60.03     & 59.05\\
Ensemble-LOB     & 65.95       & 64.72        & 65.95     & 65.23\\
Ensemble-MBO-LOB     & \textbf{66.17}       & 64.78        & \textbf{66.17}     & \textbf{65.34}\\
\midrule
\multicolumn{5}{c}{\textbf{Prediction horizon k = 100}}                 \\
\midrule
MBO-LM        & 41.27       & 34.23        & 41.27     & 35.05 \\
MBO-MLP       & 44.19       & 42.70        & 44.19     & 40.29 \\
MBO-LSTM      & 57.96       & 54.10        & 57.96     & 54.79 \\
MBO-Attention & 56.36       & 53.66        & 56.36     & 53.75 \\
\midrule
LOB-LM        & 46.19       & 43.29        & 46.19     & 41.80 \\
LOB-MLP       & 48.36       & 47.39        & 48.36     & 43.66 \\
LOB-LSTM      & 61.27       & 58.47        & 62.82     & 57.97 \\
LOB-CNN       & 61.78       & 56.91        & 61.78     & 55.40 \\
LOB-DeepLOB   & 62.82       & \textbf{60.94}        & 61.27     & 61.10\\
\midrule
Ensemble-MBO     & 56.62       & 54.85        & 56.62     & 55.48\\
Ensemble-LOB     & 63.25       & 59.41        & 63.25     & 60.56\\
Ensemble-MBO-LOB     & \textbf{63.75}       & 60.01        & \textbf{63.75}     & \textbf{61.82}\\
\bottomrule
\end{tabular}}
\label{tb:results}
\end{table}

\begin{figure}[!t]
\centering
\includegraphics[width=5.5in, height=2in]{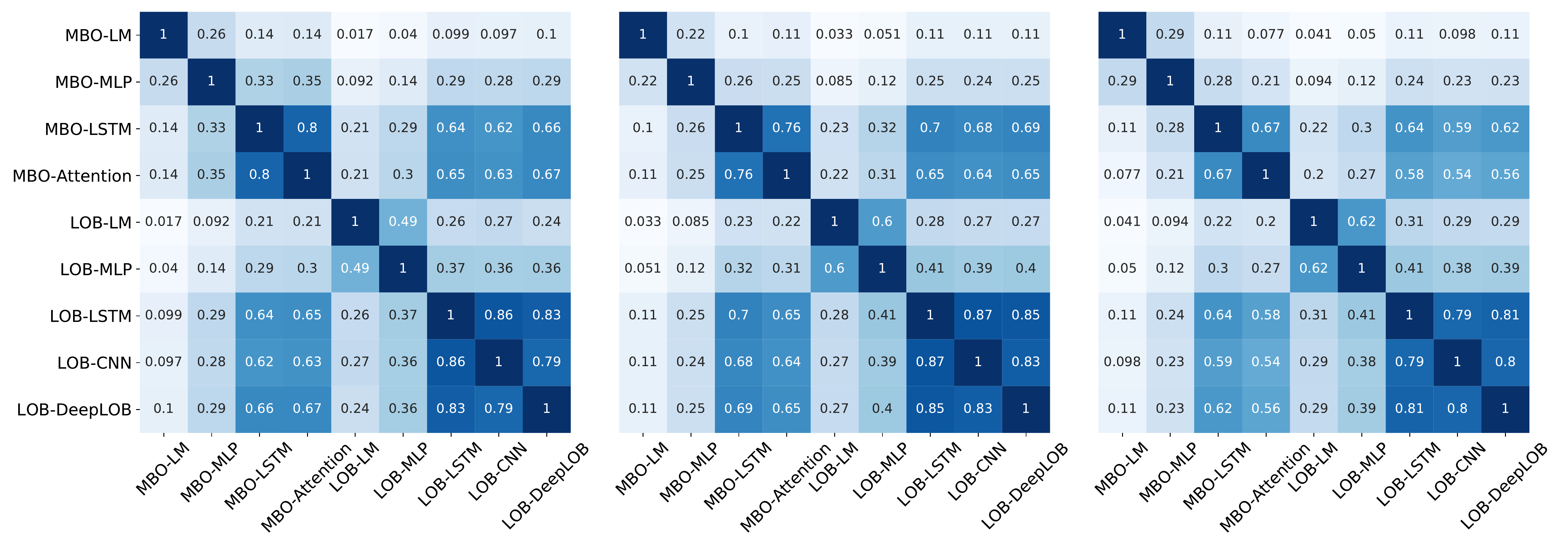}
\caption{Pearson correlation between different predictive signals for different prediction horizons ($k$). \textbf{Top:} $k=20$; \textbf{Middle:} $k=50$; \textbf{Bottom:} $k=100$.}
\label{fig:results_correlation}
\end{figure}

We observe that the models trained with LOB data are comparable, but slightly outperform the ones using MBO data. While a priori, MBO data contains more information (contents of level and trades), it is harder to model the raw messages rather than LOB snapshots when can be seen as derived or handcrafted features from the MBO data. What is encouraging is that we are able to obtain comparable performance by modelling the raw messages directly. Furthermore, if we look at the Pearson correlation between predictive signals in Figure~\ref{fig:results_correlation}, we can see that predictive signals from the MBO data are less correlated with LOB's signals. This means that we were indeed able to extract different information from the MBO data. It also suggests that a combination of two signals, from MBO and LOB data, can benefit from diversification that reduces signal variance given the low correlation. 

To verify this statement, we include three ensemble models in our experiment, where Ensemble-MBO is obtained from MBO-LSTM and MBO-Attention; Ensemble-LOB is from LOB-LSTM, LOB-CNN, and LOB-DeepLOB; and Ensemble-MBO-LOB combines Ensemble-MBO and Ensemble-LOB. A equal weighting scheme is used to construct ensemble models and we can observe that ensemble approaches, in general, improve predictive performance. In particular, Ensemble-MBO-LOB delivers the best performance, indicating the potential benefits of combining the MBO and LOB data.

Since this work aims to study MBO data, we focus on analysing results from the models trained using the MBO data. We can see that the deep learning models outperform the simple linear model, suggesting the existence of nonlinear features in financial time-series, and networks are capable of extracting such features from the raw messages in MBO data. We observe that MBO-MLP delivers inferior results compared to other networks. This is most liekly due to the structure of the MLP which has full connectivity between input and hidden units -- leading MLPs to often underperform when compared to other networks in financial applications with low signal-to-noise ratio. MBO-LSTM and MBO-Attention all have a recurrent structure with parameter sharing that enables hidden states to summarise past information and update status with current observations. Such a process filters unnecessary input components and naturally models the propagation of order flow. This observation has also been reported by \cite{lim2019enhancing, zhang2020deep, zhang2020deeppo} where they find that networks with a recurrent nature deliver better results than MLPs when modelling financial time-series.

\begin{figure}[!t]
\centering
\includegraphics[width=5.5in, height=1.1in]{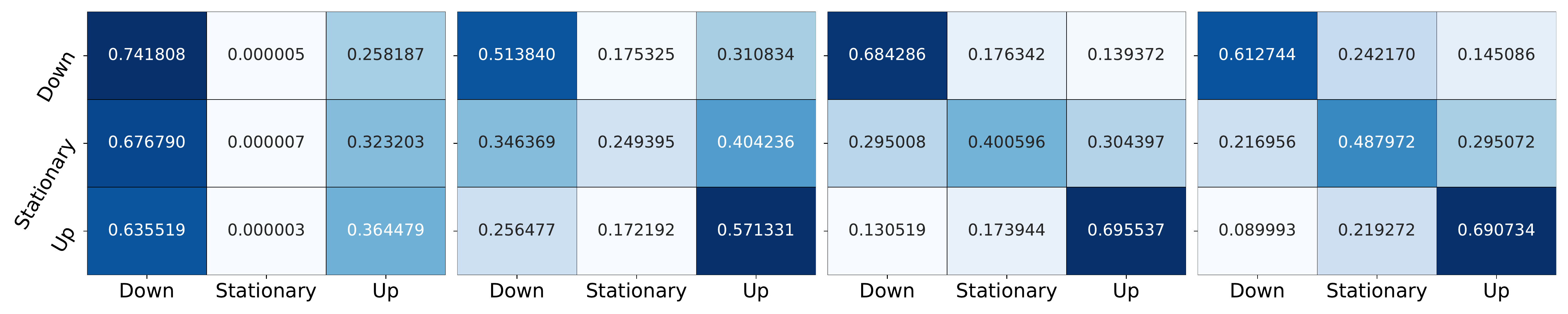}
\includegraphics[width=5.5in, height=1.1in]{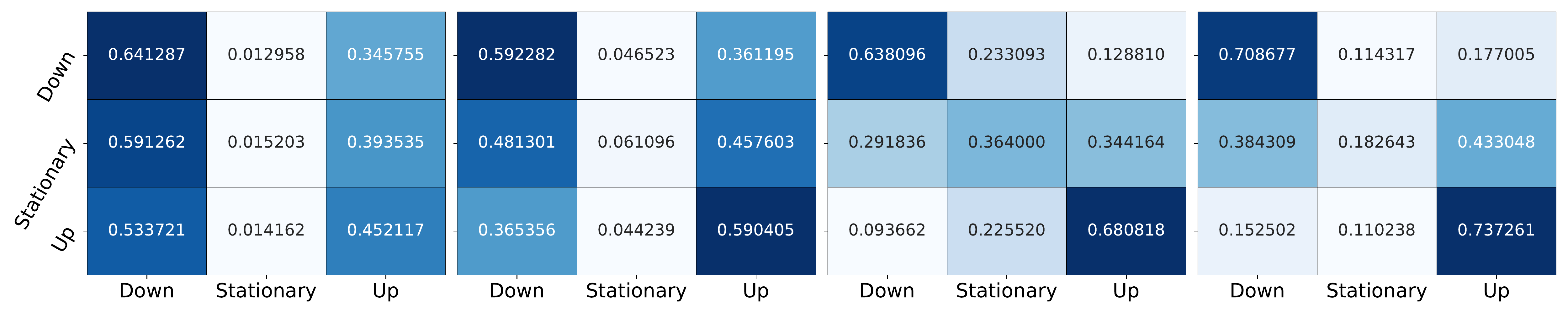}
\includegraphics[width=5.5in, height=1.1in]{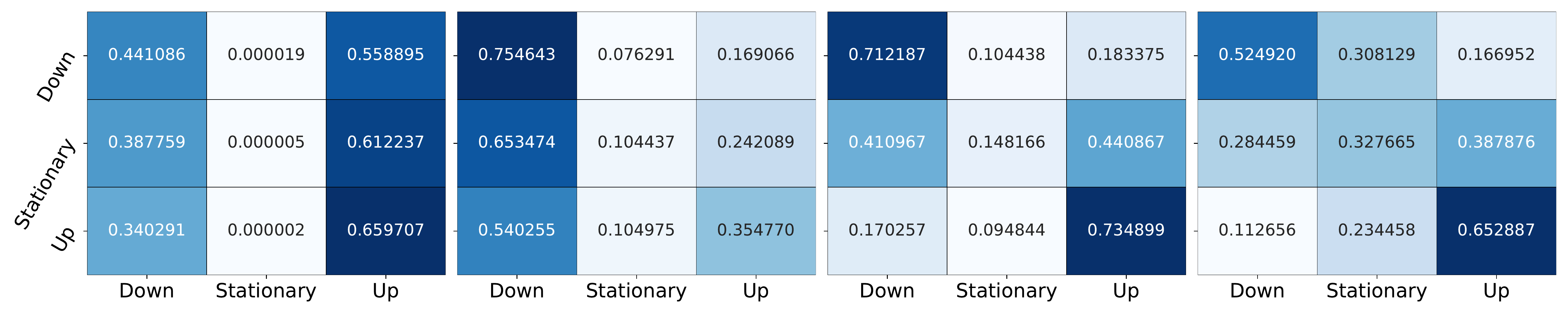}
\caption{Normalised confusion matrices for different prediction horizons ($k$). \textbf{Top:} $k=20$; \textbf{Middle:} $k=50$; \textbf{Bottom:} $k=100$. From the left to right, we display MBO-LM, MBO-MLP, MBO-LSTM and MBO-Attention.}
\label{fig:confusion_matrix}
\end{figure}

\begin{figure}[!htb]
\centering
\includegraphics[width=5.5in, height=1.6in]{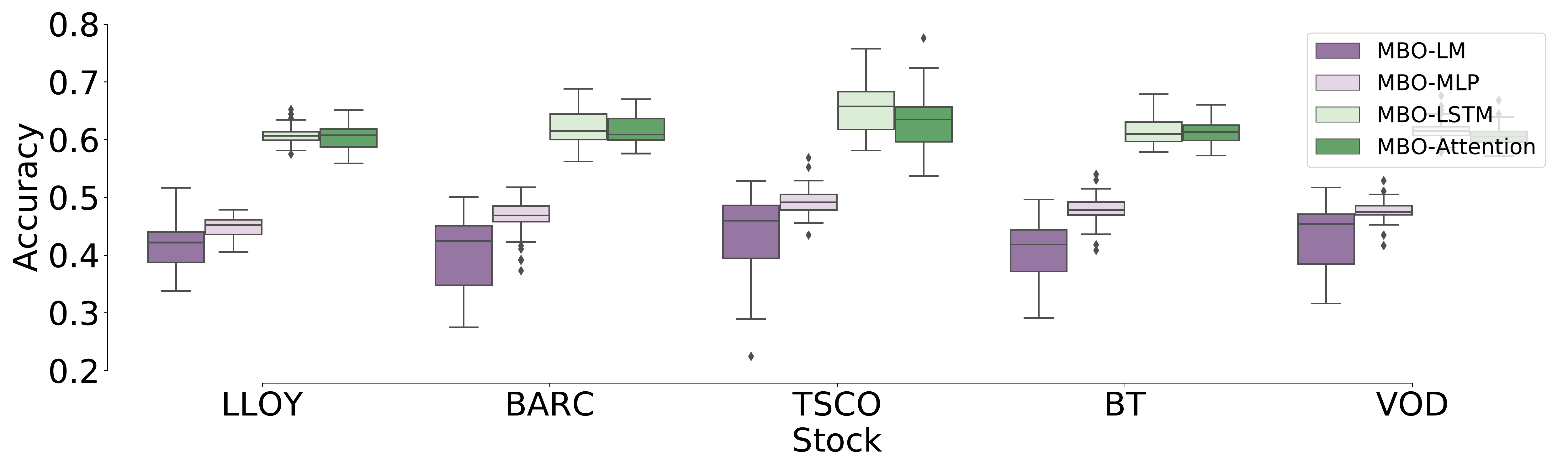}
\includegraphics[width=5.5in, height=1.6in]{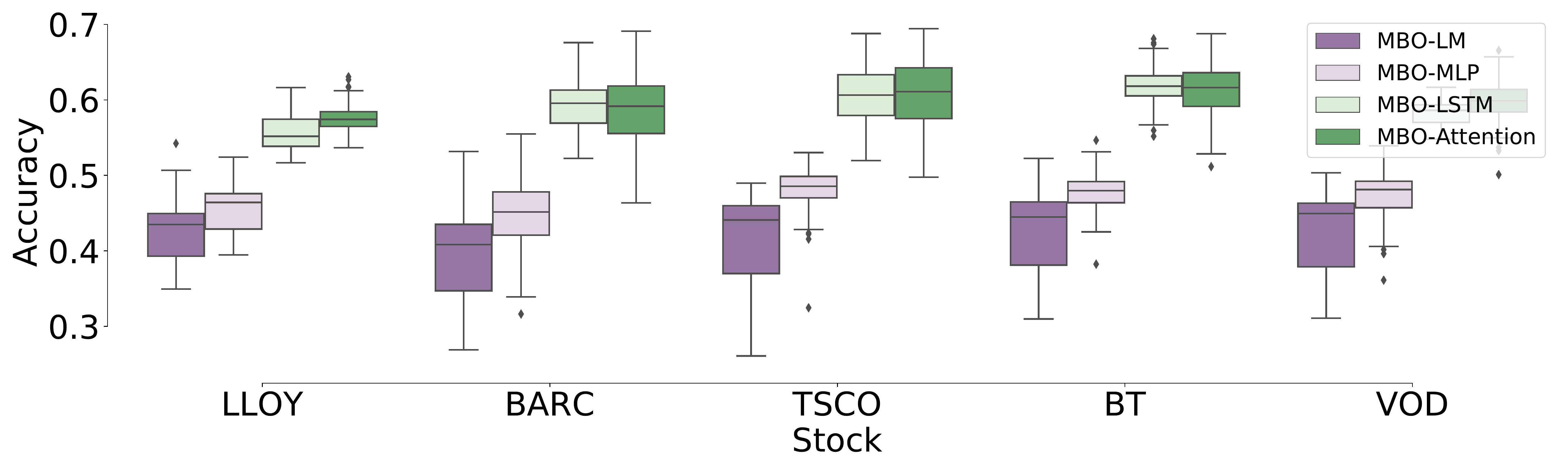}
\includegraphics[width=5.5in, height=1.6in]{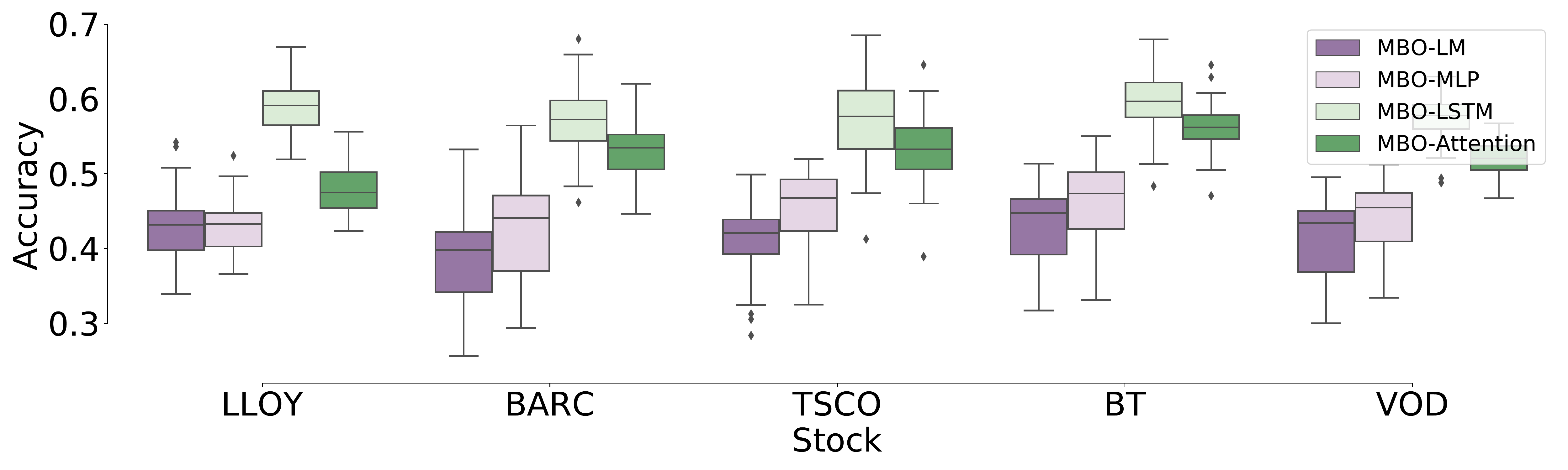}
\caption{Whisker plots of daily accuracy for different prediction horizons ($k$). \textbf{Top:} $k=20$; \textbf{Middle:} $k=50$; \textbf{Bottom:} $k=100$.}
\label{fig:daily_acc}
\end{figure}

Figure~\ref{fig:confusion_matrix} shows the normalised confusion matrices which helps to understand how models perform at predicting each label class. We calculate the accuracy score for every instrument and for each testing day to understand the consistency of our results. This is summarised in the whisker plots in Figure~\ref{fig:daily_acc}. Each point in the whisker plot represents the accuracy score for one testing day, and we make the box represents the median and interquartile range from these scores. We can see that the MBO-LM and MBO-MLP have large interquartile ranges, suggesting high variances in results, while MBO-LSTM and MBO-Attention show consistent and robust results across the entire testing period. These whisker plots allow us to understand the model performance on a daily basis to ensure the generalisability of our methods. In particular, we see that performance is consistent across the entire testing period and not focused on a few days which could be due to noise. 

\section{Conclusion}
\label{conclusion}

In this work we introduce deep learning models for Market by Order (MBO) data. To the best of our knowledge this is the first study of predictive modelling of MBO data using data-driven techniques in the academic literature. Current academic research in this direction is primarily focused on LOB data and we hope that this work helps to popularise the usage of MBO which we see as the next frontier in microstructure modelling in financial data science.  

We carefully introduce the structure of MBO data and demonstrate a specific normalisation scheme that allows model training with multiple instruments using deep learning. We consider a wide range of deep learning architectures including MLP, LSTM and attention layers. Our dataset consists of millions of sample for highly liquid instruments from the London Stock Exchange, ensuring the consistency and generalisability of our methods.  

We compare models trained using MBO and LOB data respectively. We show that we can obtain similar, but slightly inferior, performance by modelling raw MBO messages, when compared to modelling LOB data. While MBO data a priori contains more information, it is harder to model the raw messages rather than LOBs, which can be seen as derived features of the data. Importantly, we show that our models can extract additional information from the MBO data which is not captured by models trained on LOB data. This means that they can add additional value as we demonstrate in an ensemble approach that combines signals from the MBO and LOB data and delivers the best performance. 

In subsequent continuation of this work, we can apply MBO data to various financial applications including market-making or trade execution. Further, the work of \cite{briola2021deep} applies Reinforcement Learning (RL) algorithms to high-frequency trading, and it would be interesting to test the effectiveness of using MBO data within a RL framework.



\section*{Acknowledgement(s)}

The authors would like to thank members of Machine Learning Research Group at the University of Oxford for their useful comments. We are most grateful to the Oxford-Man Institute of Quantitative Finance for computing support and data access.





\appendix

\section{Complete Search Space for Hyperparameters}
\label{search_space}

\textbf{Linear Model (MBO-LM)}:
\begin{itemize}
\item Learning rate: [0.0001, 0.0005, 0.001]
\item Minibatch Size: [64, 128, 256]
\end{itemize}

\textbf{Multi-layer Perceptron (MBO-MLP)}:
\begin{itemize}
\item Number of hidden layer: [1, 2, 3]
\item Number of neurons at each layer: [32, 64, 128] 
\item Learning rate: [0.0001, 0.0005, 0.001]
\item Minibatch Size: [64, 128, 256]
\end{itemize}

\textbf{Long Short-Term Memory (MBO-LSTM)}:
\begin{itemize}
\item Number of hidden layer: [1, 2, 3]
\item Number of neurons at each layer: [32, 64, 128] 
\item Learning rate: [0.0001, 0.0005, 0.001]
\item Minibatch Size: [64, 128, 256]
\end{itemize}

\textbf{MBO-Attention}:
\begin{itemize}
\item Number of hidden layer: [1, 2, 3]
\item Number of neurons at each layer: [32, 64, 128] 
\item Learning rate: [0.0001, 0.0005, 0.001]
\item Minibatch Size: [64, 128, 256]
\end{itemize}

\section{Additional Results}
\label{add_results}

\begin{figure}[H]
\centering
\includegraphics[width=4.5in, height=1.2in]{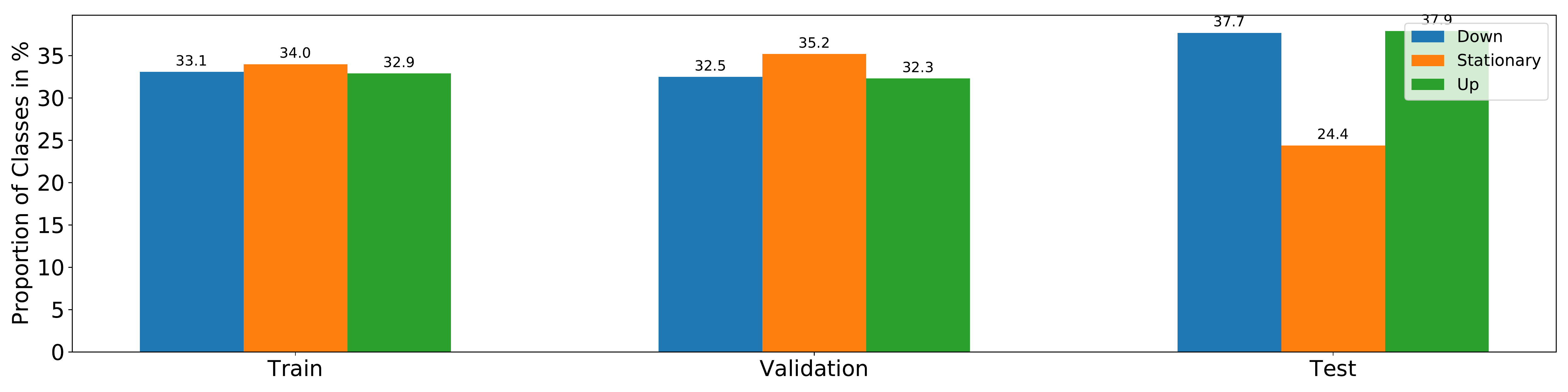}
\includegraphics[width=4.5in, height=1.2in]{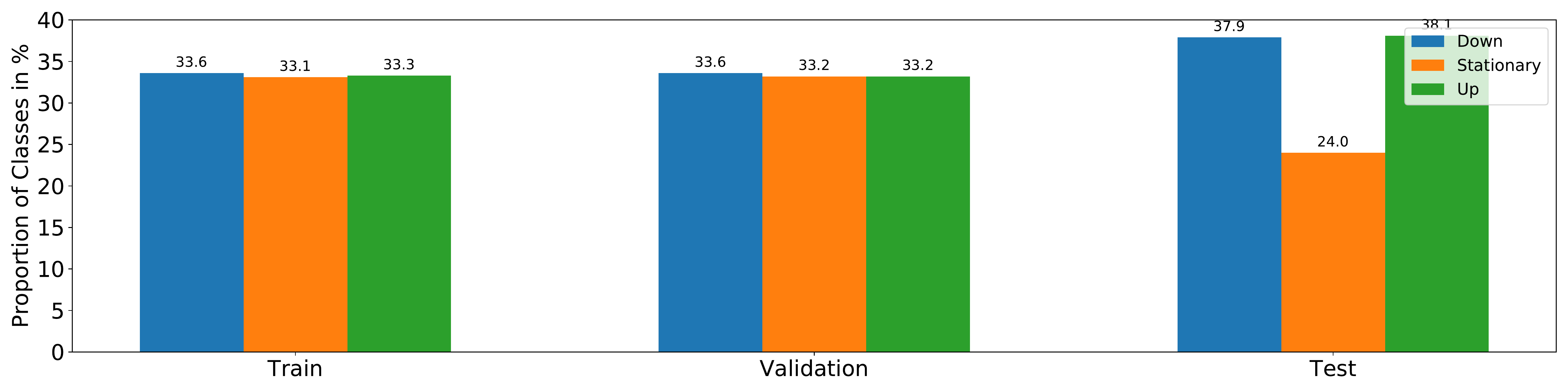}
\includegraphics[width=4.5in, height=1.2in]{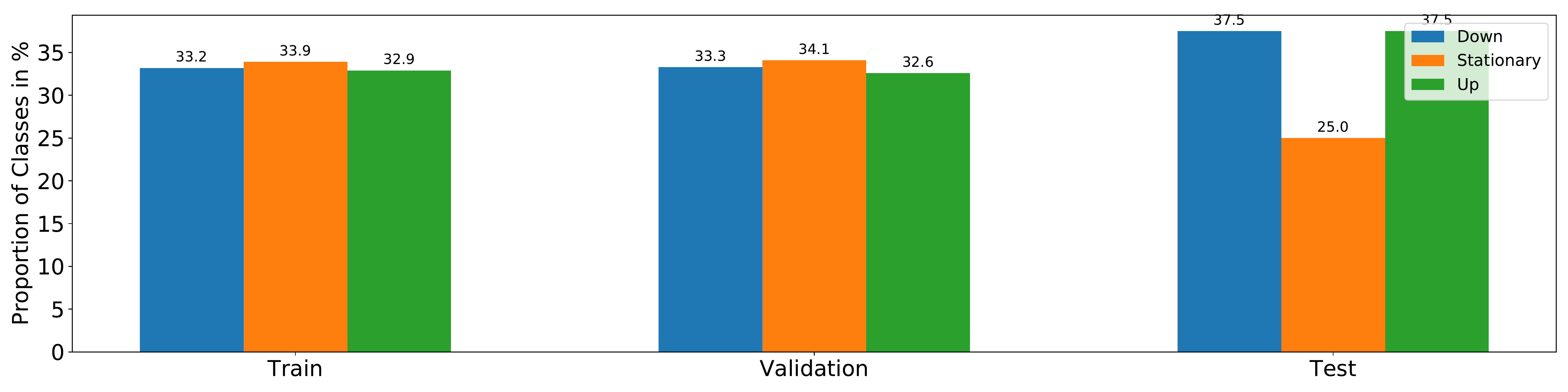}
\caption{Label class balancing for train, validation and test sets for different prediction horizons ($k$). \textbf{Top:} $k=20$; \textbf{Middle:} $k=50$; \textbf{Bottom:} $k=100$.}
\label{fig:label_class}
\end{figure}

\end{document}